\documentclass[fleqn,usenatbib]{mnras}

\usepackage[T1]{fontenc}

\DeclareRobustCommand{\VAN}[3]{#2}
\let\VANthebibliography\thebibliography
\def\thebibliography{\DeclareRobustCommand{\VAN}[3]{##3}\VANthebibliography}

\usepackage{graphicx}	
\usepackage{amsmath}	
\usepackage{amssymb}	
\usepackage{pdflscape}
\usepackage{float}

\newcommand{\kms}{\text{km} \, \text{s}^{-1}}

\newcommand{\Vsm}{V_{\rm sm}}

\newcommand{\Sint}{S_{\rm int}}
\newcommand{\rms}{\sigma_{\rm RMS}}

\newcommand{\SN}{S/N}

\newcommand{\Afr}{A_{\rm fr}}
\newcommand{\Afri}{A_{\rm fr,i}}

\newcommand{\HI}{H\,{\sc i}}

\newcommand{\dfg}{\Delta f_\text{gas}}
\newcommand{\dsSFR}{\Delta {\rm sSFR}}
\newcommand{\Msun}{{\rm M}_{\sun}}
\newcommand{\lgMstar}{\log M_{\star}}
\newcommand{\lgMstarMsun}{\log(M_{\star}/\Msun)}

\newcommand{\lgMHI}{\log(M_{\text{\HI}}/\Msun)}
\newcommand{\GF}{M_{\text{\HI}}/M_{\star}}
\newcommand{\lgGF}{\log(M_{\text{\HI}}/M_{\star})}
\newcommand{\lgsSFR}{\log{\rm sSFR}}
\newcommand{\lgsSFRyr}{\log{\rm sSFR}\, [{\rm yr}^{-1}]}

\newcommand{\cp}{\citep}
\newcommand{\ct}{\citet}

\usepackage{newtxtext,newtxmath}

\title[Global \HI\ asymmetry on the SFMS]{On the relationship between gas content, star-formation, and global \HI\ asymmetry of galaxies on the star-forming main-sequence}

\author[A. B. Watts et al.]{
Adam B. Watts,$^{1,2}$\thanks{E-mail: adam.watts@research.uwa.edu.au}
Barbara Catinella,$^{1,2}$
Luca Cortese,$^{1,2}$
Chris Power,$^{1,2}$
and Sara L. Ellison$^{3}$
\\
$^{1}$International Centre for Radio Astronomy Research, The University of Western Australia, Crawley, WA, Australia\\
$^{2}$ARC Centre of Excellence for All-Sky Astrophysics in 3 Dimensions (ASTRO3D), Australia\\
$^{3}$Department of Physics \& Astronomy, University of Victoria, Finnerty Road, Victoria, British Columbia V8P 1A1, Canada
}

\date{Accepted 2021 April 10. Received 2021 March 17}

\pubyear{2015}

\begin{document}
\label{firstpage}
\pagerange{\pageref{firstpage}--\pageref{lastpage}}
\maketitle

\begin{abstract}
Observations have revealed that disturbances in the cold neutral atomic hydrogen (\HI) in galaxies are ubiquitous, but the reasons for these disturbances remain unclear. 
While some studies suggest that asymmetries in integrated \HI\ spectra (‘global \HI\ asymmetry’) are higher in \HI-rich systems, others claim that they are preferentially found in \HI-poor galaxies.
In this work, we utilise the ALFALFA and xGASS surveys, plus a sample of post-merger galaxies, to clarify the link between global \HI\ asymmetry and the gas properties of galaxies. 
Focusing on star-forming galaxies in ALFALFA, we find that elevated global \HI\ asymmetry is not associated with a change in the \HI\ content of a galaxy, and that only the galaxies with the highest global \HI\ asymmetry show a small increase in specific star-formation rate (sSFR).
However, we show that the lack of a trend with \HI\ content is because ALFALFA misses the ‘gas-poor’ tail of the star-forming main-sequence. 
Using xGASS to obtain a sample of star-forming galaxies that is representative in both sSFR and \HI\ content, we find that global \HI\ asymmetric galaxies are typically more gas-poor than symmetric ones at fixed stellar mass, with no change in sSFR. 
Our results highlight the complexity of the connection between galaxy properties and global \HI\ asymmetry. 
This is further confirmed by the fact that even post-merger galaxies show both symmetric and asymmetric \HI\ spectra, demonstrating that merger activity does not always lead to an asymmetric global \HI\ spectrum.
\end{abstract}

\begin{keywords}
galaxies: evolution -- galaxies: ISM -- galaxies: star formation -- galaxies: kinematics and dynamics -- radio lines: galaxies
\end{keywords}



\section{Introduction}
The interstellar medium in star-forming galaxies is dominated by neutral atomic hydrogen (\HI), which exists in a thin, rotationally supported disc. 
As \HI\ is typically detectable to galactocentric radii a factor of 2-3 times higher than the bright optical component of galaxies \cp[e.g.][]{wang14}, it is a sensitive tracer of external environmental mechanisms such as ram-pressure stripping or tidal interactions. 
The impact of these mechanisms is apparent in spatially resolved \HI\ observations as asymmetric distributions of \HI\ gas or deviations from ordered kinematics \cp[e.g.][]{swaters99,chung07}, but the vast majority of \HI\ observations are only spectrally {(and not spatially)} resolved.
{Global \HI\ spectra carry} little direct spatial information about the gas, but the signature of a disturbance {nonetheless} can be encoded in the shape of the emission line profile \cp[e.g.][]{walter08,deblok20,reynolds20a}.
Quantifying and interpreting these disturbances in global \HI\ spectra, their relationship to galaxy properties, and whether they can be used to identify different evolutionary processes is vital to fully utilising the next generation of blind \HI\ surveys, which are predicted to detect over 500 000 global \HI\ spectra \cp{koribalski20}. 

{Disturbances in the \HI\ reservoirs in galaxies, measured through quantified asymmetry of global \HI\ spectra (from here `global \HI\ asymmetry'), are commonplace \cp{richter94,haynes98,espada11}, as $\sim37$ per cent of galaxies show significant global \HI\ asymmetry after correcting for noise effects \cp{watts20a}.}
{This has lead to numerous physical processes being proposed as their driving mechanisms, such as tidal interactions, gas accretion or removal, and lopsided dark matter haloes \cp[see the review by][ and references therein]{jog09}.}
{In the isolated environment, a higher rate of global \HI\ asymmetry has been found in late-type galaxies, which are typically more gas-rich  \cp{matthews98,haynes98,espada11}.}
Galaxy pairs have been found to show higher global \HI\ asymmetry compared to isolated galaxies \cp{bok19}, and \ct{ellison18} found that post-merger galaxies are systematically \HI-rich at fixed stellar mass.  
This has lead to the interpretation that \HI\ asymmetry is higher in more gas-rich systems; and that gas accretion, which is stochastic in the local Universe \cp{sancisi08}, or accretion events such as mergers, could be the driving mechanisms.

{However, recent work by \ct{reynolds20b} utilising 1167 galaxies limited to the gas-rich regime from the \HI\ Parkes All-Sky Survey \cp[HIPASS,][]{barnes01,meyer04,wong06} found no strong correlation between global \HI\ asymmetry and \HI\ content.}
{An even more extreme result was found by \ct{watts20a}, who analysed global \HI\ asymmetry in the extended GALEX Arecibo SDSS Survey \cp[xGASS,][]{catinella18} and found that global \HI\ asymmetries are actually more common in \HI-poor galaxies.}
{This apparent contradiction between the results of \ct{watts20a}, \ct{reynolds20b}, and the higher asymmetry rates of late-type isolated galaxies and mergers likely originates from the different sample selections in these studies, and the galaxy environments to which they are sensitive.}
xGASS {detects} galaxies down to \HI\ mass-fractions ($\GF$) of $\sim2$ per cent in large group to small cluster environments, while isolated late-type galaxies and mergers are typically gas-rich systems. 
This demonstrates the need to quantify the relationship between galaxy properties and global \HI\ asymmetry in systems that are not strongly affected by environment, namely gas-rich, star-forming galaxies.

{In this work, we show how previous results are not in contradiction with each other once sample selection is taken into account. }
We take advantage of the Arecibo Legacy Fast ALFA \cp[ALFALFA,][]{giovanelli05,haynes18} blind \HI\ survey, which by construction is biased toward detecting the most \HI-rich objects in its volume, and thus samples the \HI-rich regime with high statistics. 
{We also include a sample of post-merger galaxies \cp{ellison18} to explore the parameter space of systems undergoing strong gravitational interactions and the xGASS sample of \ct{watts20a} for context outside the gas-rich regime.} 
{By combining these three samples, our data are specifically designed to span a comprehensive range in \HI\ fractions and level of disturbance.}
{Utilising this, we aim to determine how the \HI\ and star-formation content of galaxies change as a function of their global \HI\ asymmetry.}

This paper is formatted as follows. 
In \S\ref{sec:data} we describe our three data-sets, and in \S\ref{sec:methods} we describe our spectrum fitting, asymmetry measurements, quality cuts, population selections, and {our method of comparing galaxy populations}. 
Our results are presented in \S\ref{sec:results}, and in \S\ref{sec:concl} we discuss them in context with the literature, and conclude.
All distance-dependant quantities in this work are computed assuming a flat $\Lambda$-Cold Dark Matter cosmology with $h=0.7$, $\Omega_{\rm M} =0.3$ and $\Omega_{\Lambda}=0.7$.

\section{Datasets} \label{sec:data}

\subsection{ALFALFA sample}
We identified the best optical counterparts in The Sloan Digital Sky Survey \cp[SDSS,][]{york00} DR7 spectroscopic catalogue \cp{abazajian09} to ALFALFA detected (code=1) sources \cp{haynes18} within 10", and selected galaxies also covered by the \textit{GALEX}-SDSS-Wise Legacy Master Catalogue \cp[GSWLC-X2,][]{salim16,salim18} within the following tolerances:
\begin{itemize}
    \item $\lgMstarMsun \geq 9$
    \item $0.02 \leq z \leq 0.05$
\end{itemize}
{The stellar mass-selection is defined to cover the same range as xGASS (see \ref{subsec:xGsamp}); and the redshift selection is defined to avoid significant contributions from peculiar motions to recessional velocity measurements below $z=0.02$, and the frequency range of the San Juan airport radar affecting ALFALFA observations above $z=0.05$.}
{In \ref{subsec:fitting}, we employ further selection cuts to make sure our results are not affected by artefacts in the data.}
We use stellar masses and star-formation rates (SFR) from the GSWLC-X2 catalogue as SFRs have been estimated using energy-balance constrained fitting of the UV/optical spectral energy distribution, plus infra-red luminosity of galaxies, as described in \ct{salim18}.
This is to avoid underestimation of the SFR derived from the centrally-targeted SDSS fibre spectra in some massive star-forming galaxies \cp[e.g.][]{cortese20}, which can host passive bulges at their centre while star-formation proceeds as normal in the disc.
The GSWLC-X2 catalogue is produced from the deepest UV data for each object in the \textit{GALEX} shallow, medium, and deep catalogues, and is suited to studies of the star-forming main-sequence where UV emission is bright. 

\subsection{xGASS sample} \label{subsec:xGsamp}
xGASS \cp{catinella18} is a stellar mass- and redshift-selected sample of 1179 galaxies in the ranges $10^{9} \leq \lgMstarMsun \leq 10^{10.5}$ and $0.01 \leq z \leq 0.05$. 
The parent sample is the overlap of the SDSS DR7 spectroscopic survey, the \textit{GALEX} Medium Imaging Survey \cp{martin05}, and ALFALFA footprints. 
Galaxies not detected by ALFALFA were re-observed with targeted Arecibo observations until detected, or until an upper limit of {$\sim2-10$ per cent in $\GF$ was reached (depending on galaxy stellar mass)}, making xGASS the deepest observations of cold gas in a representative sample of galaxies in the local Universe. 
{Stellar masses in xGASS are from the Max Planck Institute for Astrophysics (MPA)/Johns Hopkins University (JHU) value-added catalogues (VAC) based on SDSS DR7, and SFRs are calculated from \textit{GALEX} near-UV (NUV) plus Wide-field Infrared Survey Explorer \cp[WISE,][]{wright10} mid-infrared (MIR) fluxes as described in \ct{janowiecki17}, and are comparable to the GSWLC-X2 catalogue used for ALFALFA.}
In this work, we use the sub-sample of xGASS detections presented in \ct{watts20a}. 
{Briefly, from the 804 xGASS detections (formal and marginal), seven galaxies were removed due to RFI overlapping the \HI\ spectrum, and 108 \HI\ confused galaxies were removed using the confusion flags in the xGASS catalogue as the detected \HI\ is not meaningfully associated with the optical target.}
This left 689 galaxies in our xGASS sample, {which we refine in \ref{subsec:fitting}}.

\subsection{Post-merger galaxies}
We include a sample of merging galaxies identified as being in the post-coalescence phase (post-mergers) as presented and described in \ct{ellison18}. 
Briefly, the sample consists of systems with $\lgMstarMsun\geq9$ and redshift $z<0.04$ compiled from: post-mergers identified in the Galaxy Zoo project by \ct{darg10} with extra quality control by \ct{ellison13}, visual classification of SDSS galaxies by \ct{nair10}, and visual inspection of galaxies with \ct{simard11} $r$-band asymmetry $>0.05$  by \ct{ellison18}.
Post-mergers without archival \HI\ detections were re-observed with Arecibo using the same survey design as xGASS, resulting in a sample of 86 post-mergers with \HI\ detections.
We cross-matched these galaxies with GSWLC-X2 to get stellar masses and SFRs consistent with our ALFALFA sample, leaving 75 galaxies.

\section{Measurements and population selections}  \label{sec:methods}
\subsection{Spectrum fitting and asymmetry measurement} \label{subsec:fitting}
\begin{figure}
    \centering
    \includegraphics[width=0.4\textwidth]{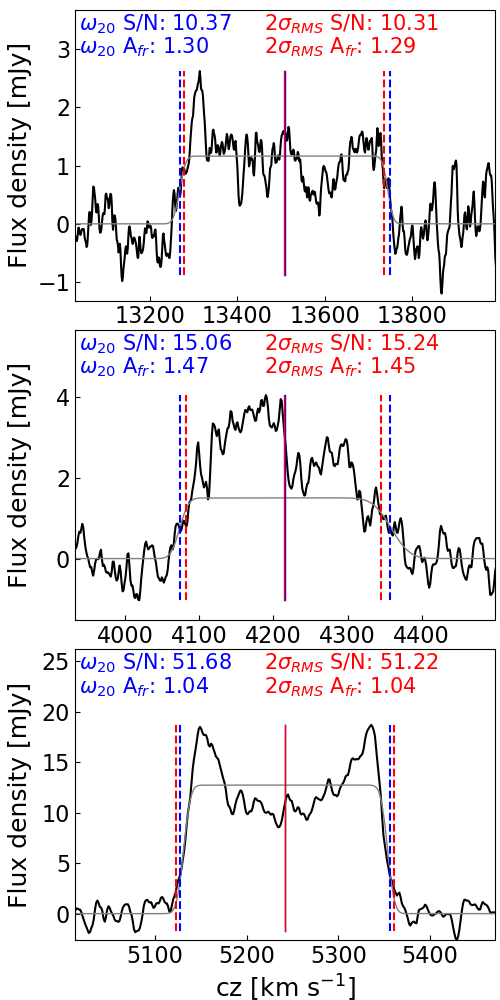}
    \caption{Comparison of measurement limits. Three xGASS spectra with $\SN \sim 10$ (top), $\SN\sim15$ (middle), and $\SN\sim51$ (bottom) with measurement limits defined at 20\% of the height of the peaks (blue) and at 2$\rms$ (red). The vertical dashed lines show the locations of $V_L$ and $V_U$, the thin-solid lines the location of $V_M$, and the coloured text corresponds to the respective $\SN$ and $\Afr$ measurements. {The thin, grey lines show the busy function fits that are focused on parameterising only the edges.}}
    \label{fig:w20rms}
\end{figure}

{As in \ct{watts20a}, we use the busy function \cp{westmeier14} to parameterise our \HI\ spectra.}
{However, here we only use the fits to identify the edges of each spectrum, without attempting to optimise the fit to the shape of the \HI\ profile.}
{Examples of busy function fits can be seen in Fig. \ref{fig:w20rms} as thin, grey lines.}
We define lower- ($V_L$) and upper- ($V_U$) velocity limits for the spectra as where the busy function fit equals twice the RMS noise ($2\rms$), measured in the signal-free part of the data, and taking the difference of these limits we measure the velocity width ($\omega$) and integrating between them we measure the integrated flux ($\Sint$).

{We computed the median and median absolute deviation ({multiplied by 1.4826 to {convert} to a standard deviation, $\sigma_{\rm MAD}$}) of the difference between our $\Sint$ measurements and the ALFALFA catalogue values, and selected any spectra with more than a {1 $\sigma_{\rm MAD}$ difference from the ALFALFA $\Sint$ value as} outliers, and all cases where at least one of the error function terms in the busy function fit was $<0.2$\footnote{This is not a universal threshold as some Gaussian-shaped spectra are fit well by this parameter value, but a value identified from visual inspection of cases where straight-edged spectra are incorrectly fit.}, as this typically means it has not meaningfully fit the edges of the spectrum.}
We manually intervened in these fits, and boxcar smoothed the spectra or fixed fit parameters as necessary to parameterise the edges.

{Using our measurements, we computed the integrated signal-to-noise ratio of each spectrum \cp[$\SN$,][]{saintonge07}}
\begin{equation} \label{eq:SN}
   \SN = \frac{\Sint/\omega}{\rms}\sqrt{\frac{1}{2}\frac{\omega}{\Vsm}},
\end{equation}
where $\Vsm$ is the final velocity resolution of the spectrum after boxcar smoothing.
To measure asymmetry we define the middle velocity $V_M = (V_L + V_U)/2$, and compute the ratio of integrated flux in each half of a spectrum
\begin{equation} \label{eq:A}
    A = \frac{
        \int_{V_{M}}^{V_{U}} S_{\rm v} {\rm dv}
        }{
        \int_{V_{L}}^{V_{M}} S_{\rm v} {\rm dv}
        },
\end{equation}
which we use to define the integrated flux ratio asymmetry parameter \cp[e.g.][]{haynes98,watts20a}
\begin{equation} \label{eq:Afr}
    \Afr = 
        \begin{cases}
            A & A \geq 1\\
            1/A & A < 1
        \end{cases}.
\end{equation}
{This definition of $\Afr$ removes the left/right-handedness inherent to the ratio $A$, such that a perfectly symmetric spectrum has $\Afr=1$ and values of $\Afr>1$ indicate deviation from symmetry.}

{To define our ALFALFA sample, we selected 2742 galaxies with $\SN\geq10$ to avoid low-quality spectra and reduce uncertainty in the measurement of $\Afr$ \cp{watts20a}}. 
{We removed a further 251 (9\%) galaxies that had  measurement limits separated by less than 20 channels (i.e., narrower than $110\ \kms$) to avoid the regime where $\Afr$ cannot be properly determined \cp{deg20}, 142 (5\%) galaxies that had missing data due to RFI removal that overlapped the \HI\ spectral line, and a further 98 (4\%) that did not have well-defined edges and could not be fit, leaving 2251 galaxies in our sample (82\%).}
Last, we assessed our ALFALFA sample for \HI\ confusion by cross-matching each galaxy to all objects with $\lgMstarMsun \geq 8.5$  in the MPA/JHU value-added catalogue within a tolerance of {2 arcmins and 200\,$\kms$}. 
{219} galaxies had at least one companion within these limits, and we removed {191} with companions bluer than the $g-r$ colour cut of \ct{zu16} used to separate blue and red sequence galaxies,
\begin{equation}
    g-r < 0.8 \,\Big(\frac{\lgMstar}{10.5}\Big)^{0.6},
\end{equation}
using the SDSS model magnitudes of each companion.
{Only galaxies with blue companions are removed, as their colour indicates that they are star-forming and thus host cold gas that could contribute non-negligible \HI\ emission to the spectrum of the target galaxy, while red companions are likely to be passive and not host significant cold gas}.
We inspected the 2032 galaxies with no identified companions and removed 76 clear cases of confusion where the companion is not part of the SDSS DR7 catalogue.
{Thus, the final number of galaxies in our ALFALFA sample is 1984\footnote{{We also tested stricter confusion limits of 3 arcmins and 300\,$\kms$, which removes ~$\sim$200 more galaxies, and found no quantitative changes to our results.}}.}

{The post-merger galaxies were fit using the same procedure, and the quality cuts reduced the sample size from 75 galaxies to 45.}
{We utilised the busy function fits to xGASS spectra from \ct{watts20a}, remeasured them using 2$\rms$ measurement limits, and applied the same quality cuts to ensure consistency between the samples, leaving 399 galaxies.}
{Both the post-merger and xGASS catalogues have already been assessed for \HI\ confusion.}
{We refer to these samples of 1984, 399, and 45 galaxies as the parent-samples for ALFALFA, xGASS, and the post-mergers, respectively.}

{Finally, we note that our velocity limits are not a `standard' choice for \HI\ studies, which typically adopt some fraction of the peak(s) in the spectrum.}
{The values of these peak(s) are not immune to elevation or suppression due to noise, particularly at low $\SN$, and we have not parameterised them with our busy function fits.}
{In Fig. \ref{fig:w20rms}, we show how our velocity limits compare to the choice of 20\% of a spectrum's peak(s), as adopted by \ct{watts20a} for xGASS, for three different $\SN$ spectra.}
{Clearly, there is no significant difference between the locations of the limits, or the resulting measurements of $\SN$ or $\Afr$ in these examples.}
{We also compared the differences in these measurements for our whole xGASS sample and found a median and $\sigma_{\rm MAD}$ of 0.017 and 0.124 for $\SN$, and -0.0016 and 0.014 for $\Afr$, respectively.}

\subsection{Star-forming main-sequence selection}

\begin{figure}
    \centering
    \includegraphics[width=0.49\textwidth]{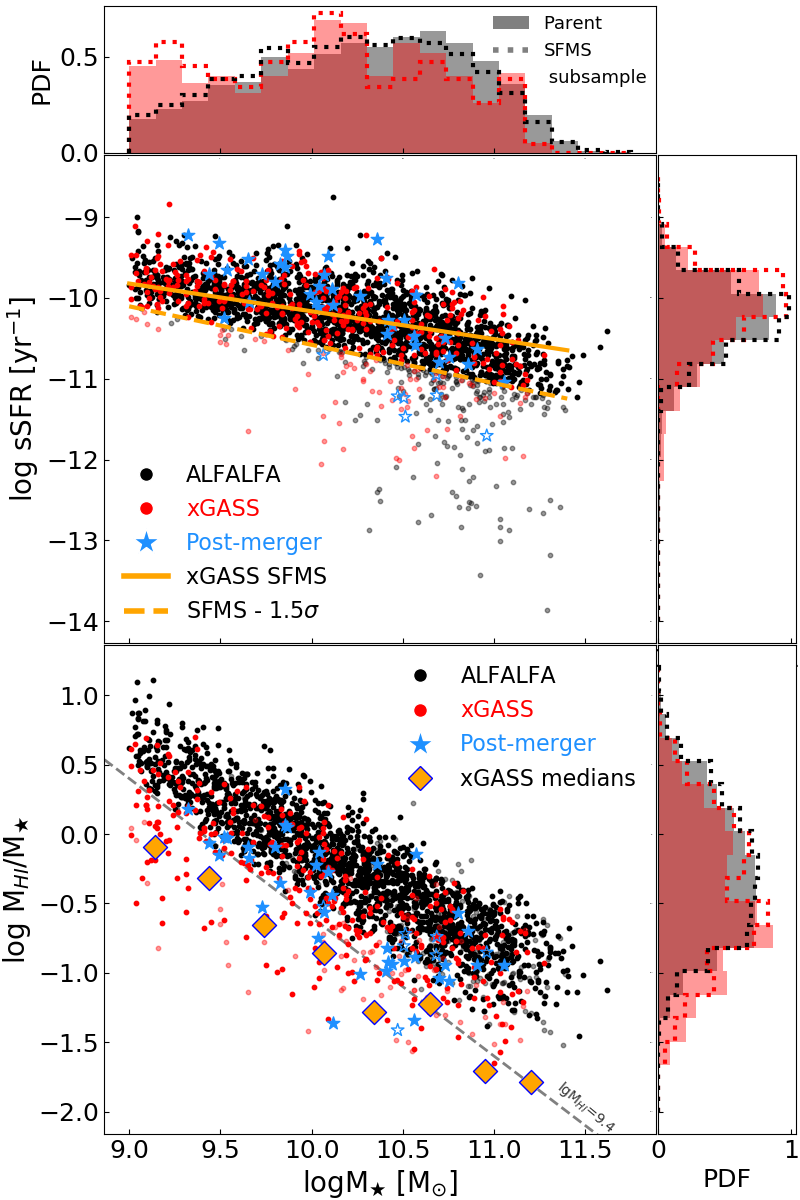}
    \caption{Scaling relations and property distributions of our samples. The top panel shows sSFR as a function of stellar mass, with the xGASS SFMS shown as a solid, orange line. The bottom panel shows the \HI\ mass fraction scaling relation with the xGASS weighted medians {\cp[from][computed including non-detections set to their upper limits]{catinella18}} shown as orange diamonds. The density-normalised probability distributions of each parameter are given on the opposite side to their respective axis labels. In both panels, ALFALFA and xGASS are shown as black and red, respectively, and the distinction between solid/light points or dotted/filled histograms is by membership to the sub-/parent- samples, where the sub-samples are separated by SFMS$-1.5\sigma$ as shown by the dashed orange line in the top panel. Post-merger galaxies are shown using filled/open, light blue stars that represent the same sub-/parent- sample membership. The dashed, grey line in the bottom panel corresponds to a constant $\lgMHI=9.4$.}
    \label{fig:samples}
\end{figure}

To study galaxies that have not undergone significant suppression of their SFR, we restrict our samples to the star-forming main-sequence (SFMS).
To do so, we use the fits by \ct{janowiecki20} to the xGASS specific SFR (${\rm sSFR} = {\rm SFR}/M_{\star}$) SFMS
\begin{equation}
    \log{\rm sSFR}_{\rm MS} \,[{\rm yr}^{-1}] = -0.344(\lgMstarMsun - 9) - 9.822,
\end{equation}
and its scatter
\begin{equation}
    \sigma_{\rm MS} [{\rm yr}^{-1}] = 0.088(\lgMstarMsun - 9) + 0.188,
\end{equation}
and select galaxies with $\lgsSFRyr \geq \log{\rm sSFR}_{\rm MS} -1.5\sigma_{\rm MS}$ from the SFMS. 
{Namely, galaxies more star-forming than $1.5 \sigma_{\rm MS}$ \emph{below} the SFMS.}
{The final number of galaxies in the SFMS sub-samples for ALFALFA, xGASS, and the post-mergers are 1784, 322, and 38, respectively, which from here we denote as  ALFALFA$_{\rm MS}$, xGASS$_{\rm MS}$, and PM$_{\rm MS}$.}

In Fig. \ref{fig:samples} we show the {stellar mass versus sSFR plane and \HI\ gas fraction scaling relation} for our xGASS, ALFALFA, and post-merger {parent-samples and SFMS sub-samples}, alongside density-normalised probability distributions of the parameters for xGASS and ALFALFA. 
We do not show the parameter distributions for the post-mergers due to their small sample size. 
The $\lgMstar$ distributions of ALFALFA and xGASS show that they cover the same $\lgMstar$ range, {with xGASS having slightly more lower mass galaxies}, and that there are no significant differences between the parent-samples and SFMS sub-samples.
In the top panel, we see that even before the selection of the SFMS, the xGASS and ALFALFA parent-samples are dominated by main-sequence galaxies. 
This is unsurprising as we can only measure asymmetry in {high $\SN$} \HI\ detections, {which are more likely to be star-forming systems.}
Both before and after the sSFR cut the $\lgsSFR$ distribution of both samples are well matched. 
Post-mergers galaxies are, as expected, preferentially star-forming systems above the SFMS due to the star-formation triggered during the merging process \cp{ellison13}.

In the bottom panel, the main difference between xGASS and ALFALFA is clear:  ALFALFA almost exclusively detects galaxies above the xGASS median\footnote{weighted medians computed including upper limits.} \HI\ mass fraction scaling relation, shown as orange diamonds. 
The xGASS parent- and SFMS sub-samples show a similar bias toward galaxies above the median scaling relation, due to being detection-selected, but the targeted survey design of xGASS means that there are also galaxies below the medians. 
Selecting the SFMS causes a slight shift toward higher \HI\ fractions, as gas content and star-formation are correlated, but the effect is small for both samples. 
{The PM$_{\rm MS}$} sample occupies the same region of the parameter space as xGASS as they have the same survey design and are also detection-selected.
These $\lgGF$ distributions demonstrate the advantage of analysing these three samples together: ALFALFA samples the gas-rich regime with high statistics, xGASS extends $0.5-1$ dex lower in \HI\ mass-fraction than ALFALFA, and the post-mergers provide strongly interacting, star-forming galaxies with similar coverage as xGASS.
{However, it also highlights how a SFMS-selected sample does \emph{not} translate to a clear selection in galaxy \HI\ mass fraction.}
{ALFALFA is not representative of the \HI\ fractions of galaxies within the SFMS, missing the `gas-poor' tail of star-forming galaxies that is revealed by xGASS, and this will limit our ability to use ALFALFA to look for trends between \HI\ asymmetry and gas content.}


\subsection{Asymmetric and symmetric galaxies} \label{subsec:popsel}
The uncertainty in an $\Afr$ measurement is closely linked to the $\SN$ of a spectrum \cp{watts20a}, and this must be taken into account when selecting and comparing different samples, particularly if they have different $\SN$ distributions.
We briefly describe how we model and account for this effect here, but refer the reader to \ct{watts20a} for a more detailed description. 
{We also acknowledge that there are other sources of uncertainty on $\Afr$ measurements, as mentioned in \ct{watts20a}, but these are minimised by our $\SN$ cut and removal of confused systems, and we do not expect them to impact our results.}

A toy model is used to create noiseless, model \HI\ spectra with intrinsic asymmetries ($\Afri$) of  $\Afri=1$, 1.1, and 1.25, {corresponding to symmetric spectra, asymmetric spectra \cp{haynes98,watts20a}, and a higher value to model strongly-asymmetric spectra.} 
Mock observations are created by degrading the model spectra to a desired $\SN$ by adding Gaussian random noise, and 10$^4$ noise realisations are created for $\SN$ values in the range $\SN=[5,100]$ in steps of $\Delta\SN=1$, {to properly model marginal detections below our $\SN$ cut to high $\SN$ spectra.} 
The `observed' $\SN$ and $\Afr$ of each mock spectrum is calculated using measurement limits defined on the noiseless model spectrum {by treating it as a `fit' to the mock spectra, as the straight edges of \HI\ spectra are typically well-defined and not significantly impacted by noise.} 

\begin{figure}
    \centering
    \includegraphics[width=0.5\textwidth]{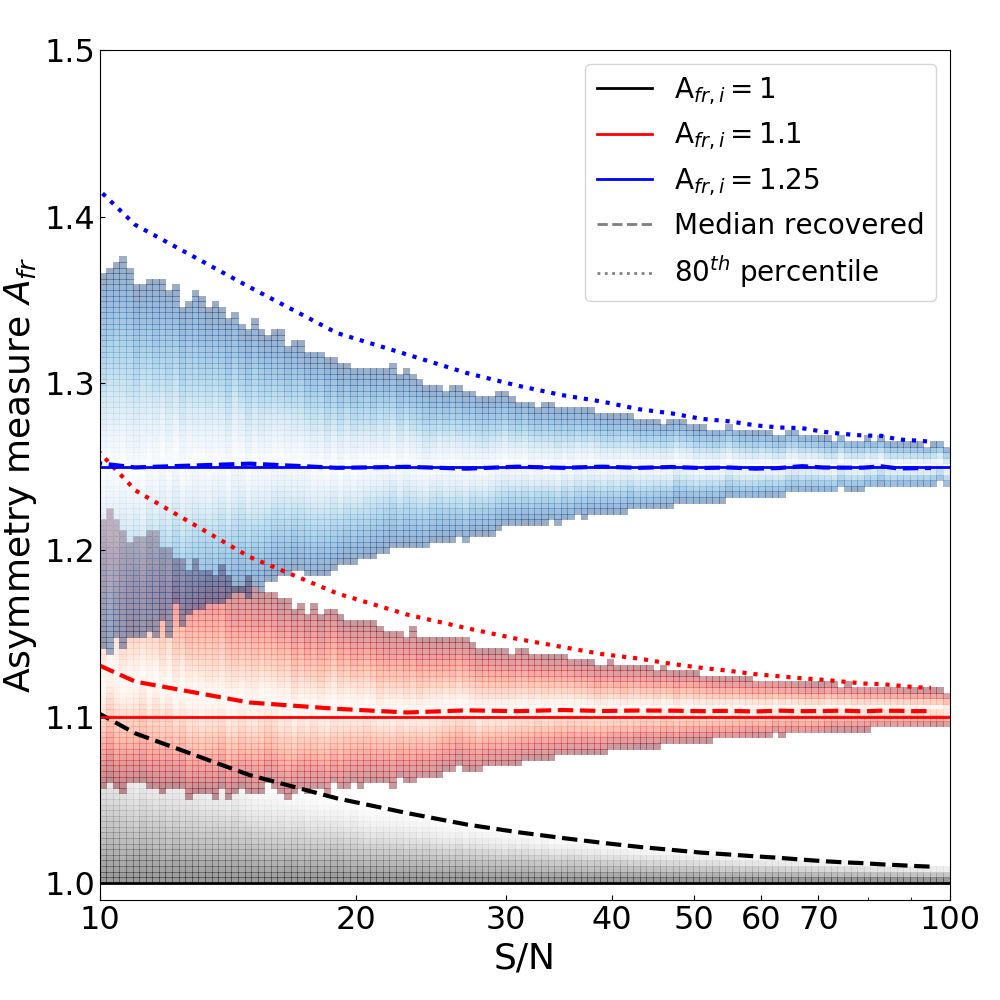}
    \caption{{The distribution of recovered $\Afr$ measurements as a function of $\SN$. The background density plots show the normalised probability distribution of $\Afr$ measurements in bins of $\Delta \log(\SN)=0.008$ for our model spectra with $\Afri=1$ (black), $\Afri=1.1$ (red) and $\Afri=1.25$ (blue). The density maps are truncated at their $25^{\rm th}$ and $75^{\rm th}$ percentiles for $\Afri=1.1$ and $\Afri=1.25$, and the $50^{\rm th}$ for $\Afri=1$ as it is symmetric about $\Afr=1$. The value of $\Afri$ for each model is shown with a solid line, and the median recovered value in bins of $\Delta \SN=4$ are shown as dashed lines. The $80^{\rm th}$ percentiles of the $\Afri=1.1$ and 1.25 models, used to select our asymmetric and strong-asymmetry populations, respectively, are shown with dotted lines.}}
    \label{fig:SN_Afr_density}
\end{figure}

{In Fig. \ref{fig:SN_Afr_density}, we show the distribution of recovered $\Afr$ measurements as a function of $\SN$ for our three modelled $\Afri$ values.}
{The background density plots show the normalised probability distribution function of recovered $\Afr$ measurements in bins of $\Delta \log(\SN)=0.008$, sampled in bins of $\Delta\Afr=0.0034$, and truncated at their $25^{\rm th}$ and $75^{\rm th}$ percentiles (the $50^{\rm th}$ for the $\Afri=1$ model, as it is reflected about $\Afr=1$).}
{The intrinsic asymmetry of each model and the median recovered $\Afr$ in bins of $\Delta\SN=4$ are shown with black ($\Afri=1$), red ($\Afri=1.1$), and blue ($\Afri=1.25$), solid and dashed lines, respectively.}
{Clearly, the uncertainty in an $\Afr$ measurement increases as a $\SN$ decreases, and at low $\SN$ galaxies preferentially scatter toward \emph{higher} $\Afr$ values due to the distributions being bounded by $\Afr=1$.}
{This is visible in the distribution of recovered $\Afr$ measurements for the $\Afri=1$ model, as the median recovered $\Afr$ is $>1$ for all $\Afr$; and in the $\Afri=1.1$ model by the deviation of its median to $\Afr>1.1$, and the upward compression of the low-$\Afr$ side of its distribution.}

We quantify this scatter in $\Afr$ measurements due to noise at a given $\SN$ by computing the percentiles of the $\Afr$ distribution for all mock spectra in bins of $\Delta\SN=4$.
These percentiles are well described by an inverse power law of the form 
\begin{equation}
    P{\rm XX}(\SN,\Afri) = \frac{1}{a(\SN - b)} + \Afri,
\end{equation}
where the constants $a$ and $b$ depend on the desired percentile `$P$XX'. 
We fit these percentiles to parameterise $\Afr$ as a function of $\SN$ for a given $\Afri$.
It is important to mention that these percentiles are almost completely insensitive to the degree of smoothing and shape of a \HI\ spectrum, so long as they are parameterised as a function of $\SN$ as defined in eq. \ref{eq:SN} \cp{watts20a}.

{Utilising this model, we select asymmetric and symmetric populations from each SFMS sub-sample using the same thresholds as \ct{watts20a}, with an additional `strong-asymmetry' population to investigate the properties of the most asymmetric galaxies}.
\begin{itemize}
    \item {The symmetric population is selected as $\Afr \leq P50(\SN,\Afri =1)$ (i.e. the median), as 50 per cent of intrinsically symmetric spectra will have $\Afr$ below this percentile.} 
    \item {The asymmetric population is defined as $\Afr \geq P80(\SN,\Afri=1.1)$, i.e. galaxies that have at least 10 per cent intrinsic asymmetry to at least an 80\% confidence level, considering noise.}
    \item {The strong-asymmetry population is defined as $\Afr \geq P80(\SN,\Afri=1.25)$, namely at least 25 per cent intrinsic asymmetry with 80\% confidence. }
\end{itemize}
{These selections are visible in Fig. \ref{fig:SN_Afr_density} as the black dashed (symmetric), dotted red (asymmetric), and dotted blue (strong-asymmetry) lines.}
There will be some contamination of the symmetric population, particularly at lower $\SN$, {but we must compromise between sample size and purity,} and the deviation from symmetry shown by these galaxies is small compared to the effect of noise, so we do not expect them to affect our results.
{Fig. \ref{fig:SN_Afr_density} also demonstrates why it is necessary to use these percentiles to select our populations, as they increase the confidence in the separation of intrinsically symmetric and asymmetric galaxies.}
{At $\SN=10$, a galaxy with $\Afr=1.1$ has a 50 per cent chance of having $\Afri=1$, namely being intrinsically symmetric.}
{Similarly, a galaxy with $\Afr=1.25$ has a 20 per cent chance of having $\Afr=1.1$, namely it is asymmetric, but not strongly-asymmetric, as might be inferred if the effect of noise was not accounted for.}
{Thus, the} number of galaxies in the asymmetric and symmetric populations, respectively, are 121 and 87 for xGASS$_{\rm MS}$, {12 and 8 for PM$_{\rm MS}$}, and 541 and 431 for ALFALFA$_{\rm MS}$.
{The strong-asymmetry population is only defined for ALFALFA$_{\rm MS}$, and consists of 168 galaxies.}

\subsection{Offset parameters} \label{subsec:offsets}
We quantify the difference in the \HI\ and star-formation properties of asymmetric galaxies compared to symmetric ones with a matched-galaxy offset analysis \cp[e.g.][]{ellison18,watts20a}, {using only galaxies within the SFMS}.
{Each asymmetric galaxy is matched to all symmetric galaxies within 0.1 dex in $\lgMstarMsun$ and $\log\SN$, and if less than five symmetric galaxies are matched then we expand these tolerances {by 0.1 and 0.05 dex, respectively}, until at least 5 matches are found.}
{Typically $>80\%$ of galaxies find sufficient matches without needing to increase the tolerances in ALFALFA$_{\rm MS}$, and with only one expansion in xGASS$_{\rm MS}$ due to the smaller sample size.}
The $\SN$ match is included because lower $\SN$ galaxies typically have higher $\Afr$ due to the effect of noise and, in xGASS, there is a correlation where higher $\SN$ galaxies typically have higher \HI\ content.
{This correlation is also present in xGASS$_{\rm MS}$, and if left uncorrected it could bias us toward inferring higher $\Afr$ galaxies are more gas-poor.}
{This correlation is not present in ALFALFA$_{\rm MS}$ and we note that removing the $\SN$ matching does not change our results, but we keep it for consistency between analyses and with previous work.}
Last, we allow galaxies with $\SN\geq40$ to match to other galaxies with $\SN\geq40$ regardless of $\SN$, as the effect of noise on $\Afr$ is small above this value, {and to avoid discarding the fewer high $\SN$ spectra}. 

The \HI\ mass fraction ($\dfg$) and sSFR ($\dsSFR$) offsets are then defined as the logarithmic difference between the value for the asymmetric galaxy, and the median of the matched symmetric galaxies:
\begin{align}
    \dfg &= \lgGF_{\rm asym} - {\rm med}[ \lgGF_{\rm sym, match}] \\
    \dsSFR &= (\lgsSFR)_{\rm asym} - {\rm med}[ (\lgsSFR)_{\rm sym, match}].
\end{align}
In this way, we compare the properties of asymmetric galaxies to what would be expected for a symmetric galaxy at fixed $M_{\star}$ and $\SN$. 
We also compute these offsets for each symmetric galaxy using the same process, to inform us about the distribution of the symmetric population.

\section{Results} \label{sec:results}

\begin{table}
    \centering
    \caption{Median offset parameters and two-sample Kolomogorov-Smirnov test $D$- and $p$- values for the populations presented in \S\ref{subsec:AA} and \S\ref{subsec:xGASS}. }
    \label{tab:offsets_KS}
    \begin{tabular}{l c c c c} \hline
        Sample & Offset & Median & $D$-value & $p$-value \\
        \hline \hline
        AA$_{\rm MS}$ asymmetric & $\dfg$& $0.01\pm0.01 $ &0.05 & 0.45 \\
         & $\dsSFR$ & $0.04\pm0.02 $ & 0.09&0.03 \\
        AA$_{\rm MS}$ strong asymm & $\dfg$&  $0.05\pm0.02 $ & 0.11 & 0.11 \\
         &$\dsSFR$ & $0.12\pm0.03 $ & 0.18& $9\times 10^{-4}$\\
        xGASS$_{\rm MS}$ asymmetric & $\dfg$& $-0.13\pm0.03$ &0.22 &0.03 \\
         & $\dsSFR $& $0.03\pm0.04 $ & 0.15 &0.22 \\
     \hline
    \end{tabular}
\end{table}


\subsection{Global \HI\ asymmetries on the SFMS}  \label{subsec:AA}

\begin{landscape}
\begin{figure}
    \includegraphics[width=1.3\textwidth]{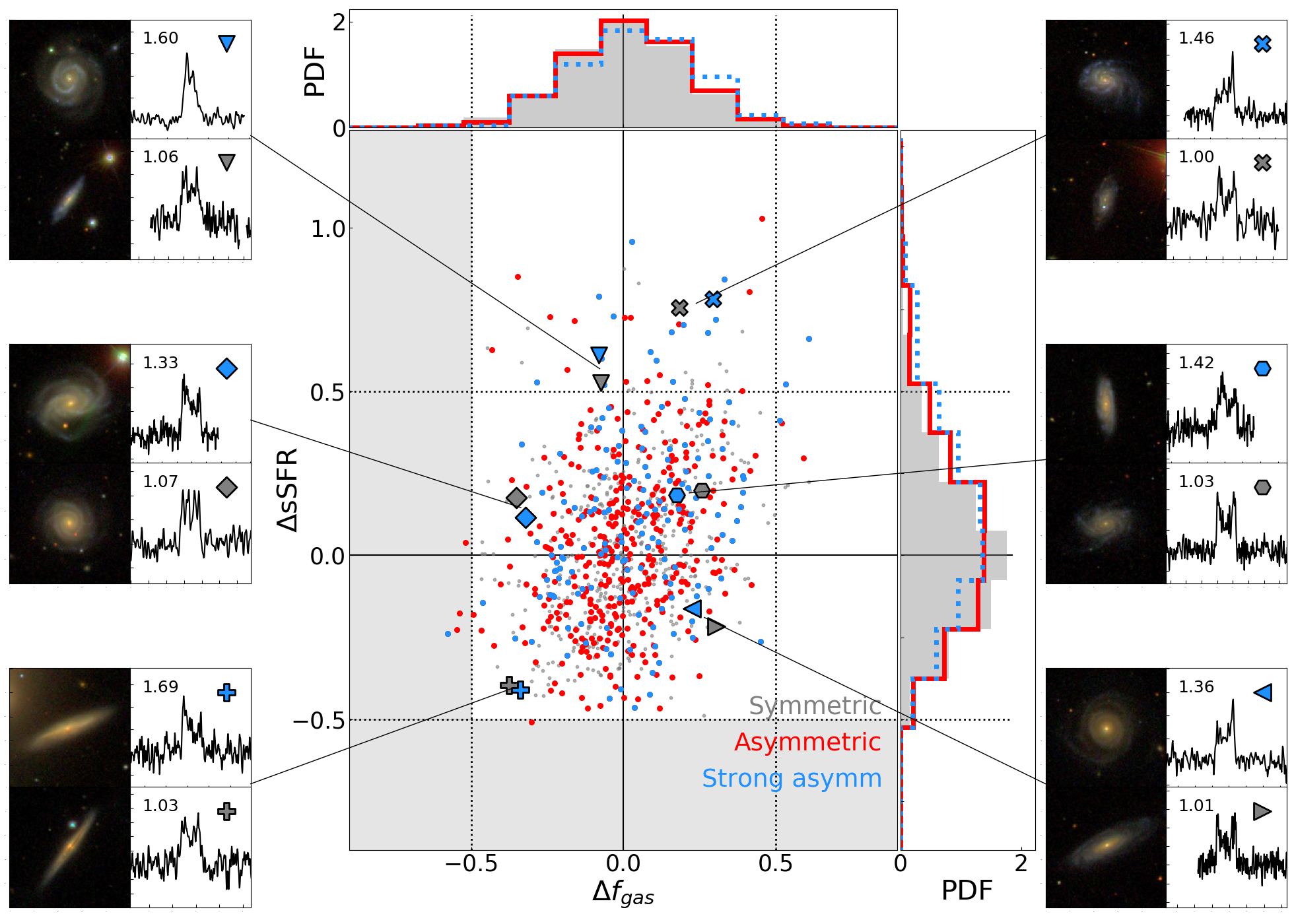}
    \caption{$\dfg - \dsSFR$ parameter space for ALFALFA$_{\rm MS}$ galaxies. The central panel shows the location of galaxies in the parameter space, and the density-normalised probability distribution of each offset parameter is shown opposite to their respective axis label. Symmetric galaxies are shown with grey points and filled histograms, asymmetric galaxies with red points and red, open histograms, and the subset of the asymmetric population with strong asymmetry as light blue points and dotted, open histograms. Horizontal and vertical black, dotted lines correspond to $\pm0.5$ for each offset parameter, and the grey, shaded regions denote offsets of $<-0.5$ {that are sparsely populated due to our sample selection criteria}. SDSS cutouts (1.5 arcmin square) and global \HI\ spectra for examples of asymmetric and symmetric galaxies are shown to either side of the main panel, and the $\Afr$ value of each spectrum is given to their top left. {The example galaxies are paired by their similar offset values, as indicated by the thin, black line connecting each pair to their location in the parameter space, and the large, bold markers in the parameter space that match the symbol next to their spectrum.}}
    \label{fig:delta_AA}
\end{figure}
\end{landscape}

{We first focus on the properties of galaxies in the ALFALFA$_{\rm MS}$ sample.}
{In Fig. \ref{fig:delta_AA} we show the $\dfg-\dsSFR$ parameter space for the ALFALFA$_{\rm MS}$ sample, where the asymmetric and strongly asymmetric samples are matched to the symmetric sample, and the symmetric sample matched to itself.} 
{We show thin, black lines at $\dfg=\dsSFR=0$ to provide a visual centre of the parameter space, and dotted, black lines at $\pm0.5$ to demonstrate the extent of the scatter.}
{To the top and right of the central panel, we show density-normalised probability distributions of each offset parameter and extensions of the $\dfg=\dsSFR=0$ and $\pm0.5$ lines.}
{Surrounding the main panel, we show  SDSS cutouts and global \HI\ spectra of example (strongly) asymmetric and symmetric galaxies, to show the diversity of optical properties of galaxies in the parameter space.}
{These example galaxies are paired according to their similar $\dfg$ and $\dsSFR$, denoted by the large, bordered, markers that are shown both in the main panel and inset beside their spectrum, with a thin, black line connecting each pair to its location in the parameter space.}

Clearly, the bulk of the symmetric and asymmetric populations are contained within $\dfg = \dsSFR = \pm0.5$.
The only clear deviation outside this scatter is for $\dsSFR>0.5$, and this part of the parameter space is occupied by both asymmetric and symmetric galaxies. 
{This is also reflected in the distributions of each offset parameter.}
{The $\dfg$ distributions are centred on zero with similar widths, while the asymmetric galaxies show slightly higher $\dsSFR$ values.}
{To provide quantitative comparisons, we calculate the median offsets of the asymmetric population with uncertainty derived from 10$^4$ bootstraps and perform a two-sample Kolmogorov-Smirnov test (KS-test) between the asymmetric and symmetric populations.}
{These medians, and the KS-test $D$-statistics and $p$-values are listed in Table \ref{tab:offsets_KS}.}
{The median offsets for the asymmetric population are small, $\dfg = 0.01 \pm 0.01 $ with $p = 0.45$ and $\dsSFR = 0.04\pm 0.02$ with $p=0.03$, though the KS-test result indicates that the $\dsSFR$ distribution of asymmetric galaxies is different to that of symmetric ones.}

{It is important to note that the} lack of galaxies with $\dsSFR<-0.5$ and $\dfg<-0.5$, shown as the grey shaded regions in the main panel of Fig. \ref{fig:delta_AA}, {is simply due to the fact that ALFALFA does not reach lower (relative) gas fractions and our sample is only representative of the SFMS}.
However, despite restricting our sample to the most gas-rich and star-forming systems, {we can already conclude that global \HI\ asymmetries are not preferentially found in gas-rich or highly star-forming objects.}

In Fig. \ref{fig:delta_AA} we also highlight the subset of asymmetric galaxies with strong asymmetry, to investigate the properties of the most disturbed systems. 
{This population spans the same $\dfg$ and $\dsSFR$ ranges as the asymmetric and symmetric populations, and the offset distributions show little difference from the asymmetric population aside from a small increase in positive offset values.}
The median offsets for this population are $\dfg=0.05 \pm0.02$ with $p=0.11$ and $\dsSFR = 0.12 \pm 0.03$ with $p=9\times10^{-4}$, indicating that {a significant offset is only present in sSFR, not \HI\ fraction.}
{This suggests that greater $\Afr$ is associated with greater $\dsSFR$.}
{{To further confirm this}, we performed the same analysis on the subset of the asymmetric population after \emph{removing} the strong asymmetry galaxies.}
{We find that this remaining subset of 373 galaxies has a median $\dsSFR=0.02\pm0.02$, namely consistent with no elevation in $\dsSFR$, and the KS-test $p=0.45$ cannot determine that the $\dsSFR$ distribution is different to that of symmetric galaxies.}
{Thus, the difference in $\dsSFR$ between symmetric and asymmetric galaxies is primarily driven by galaxies with the highest $\Afr$.}

The example galaxies surrounding Fig. \ref{fig:delta_AA} highlight the diversity in the optical properties of symmetric and strongly asymmetric galaxies across the parameter space. 
Both populations show signatures of optical disturbances, blue star-forming discs, and apparent prominence of the bulge. 
{There is no clear distinction in the optical morphologies of galaxies whether they are classified as symmetric or asymmetric, and this diversity is present across the parameter space.}
{Thus, we emphasise here the main result of this work.}
{\emph{Any physical process(es) driving \HI\ disturbances in the gas-rich regime on the SFMS are only evident in the highest $\Afr$ systems, and only appear to impact the star-formation properties of galaxies with no measurable impact on their \HI\ content.}}

\begin{figure*}
    \centering
    \includegraphics[width=0.9\textwidth]{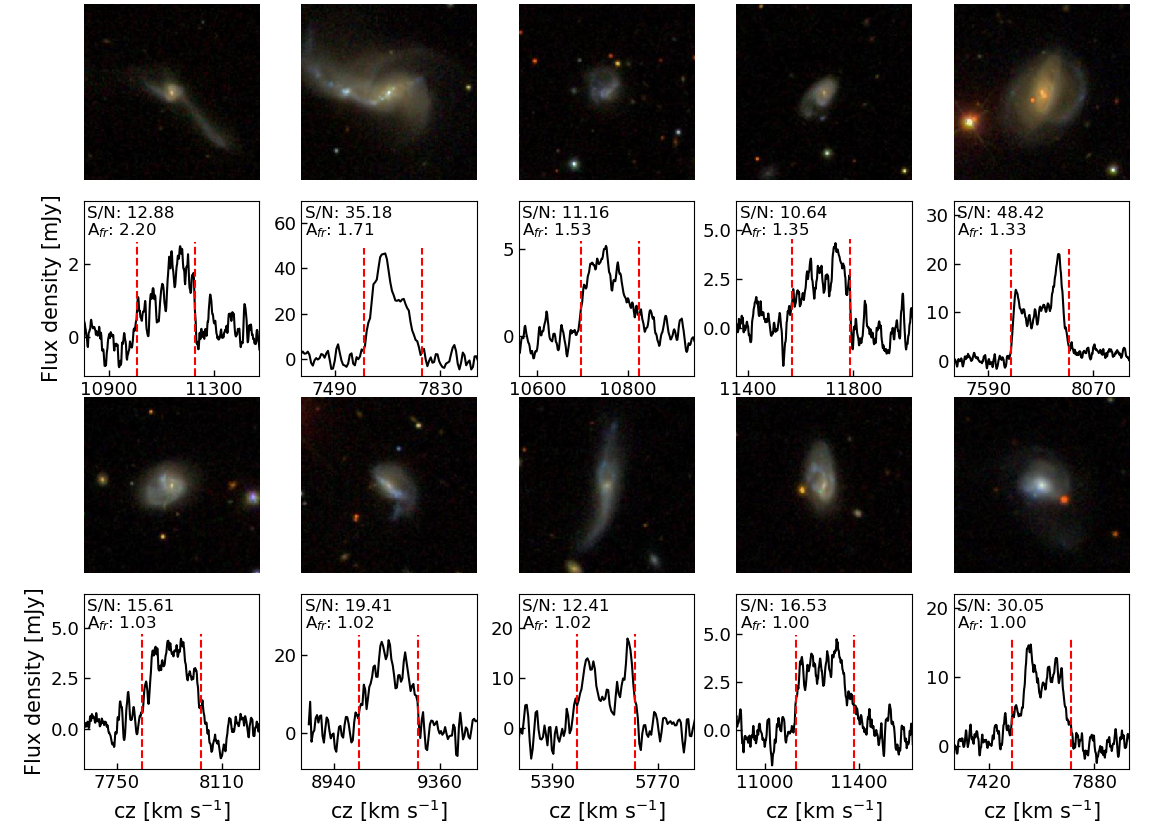}
    \caption{SDSS cutouts (1.5 arcmin square) and global \HI\ spectra of the five highest-$\Afr$ (top row) and five most symmetric (bottom row) PM$_{\rm MS}$ galaxies. The $\SN$ and $\Afr$ values are given in the top left of each spectrum.}
    \label{fig:PMexamples}
\end{figure*}

{The complexity of the link (or lack thereof) between \HI\ asymmetry and other galaxy properties is further highlighted by our sample of post-merger galaxies.}
{Our small sample size does not permit any statistical comparisons, so in Fig. \ref{fig:PMexamples} we show the five highest $\Afr$ and the five most \HI\ symmetric PM$_{\rm MS}$ galaxies.}
{The fact that PM$_{\rm MS}$ galaxies span {the same range of asymmetries -- from almost perfectly symmetric to highly asymmetric --} demonstrates that the presence of optical disturbance in a galaxy does not necessarily imply elevated \HI\ asymmetry, and vice-versa.}
{While merger activity can enhance global \HI\ asymmetry on a statistical level \cp[e.g.][]{bok19}, projection effects \cp[e.g.][]{deg20} or the settling of the gas back into regular rotation in advanced merging stages \cp[e.g.][]{manthey08,schiminovich13} could be responsible for symmetric \HI\ spectra, even when signatures of the merger remain in optical images, or stellar or ionised gas kinematics \cp[e.g.][]{feng20,nevin21}}

\subsection{Comparison with xGASS} \label{subsec:xGASS}

\begin{figure}
    \centering
    \includegraphics[width=0.5\textwidth]{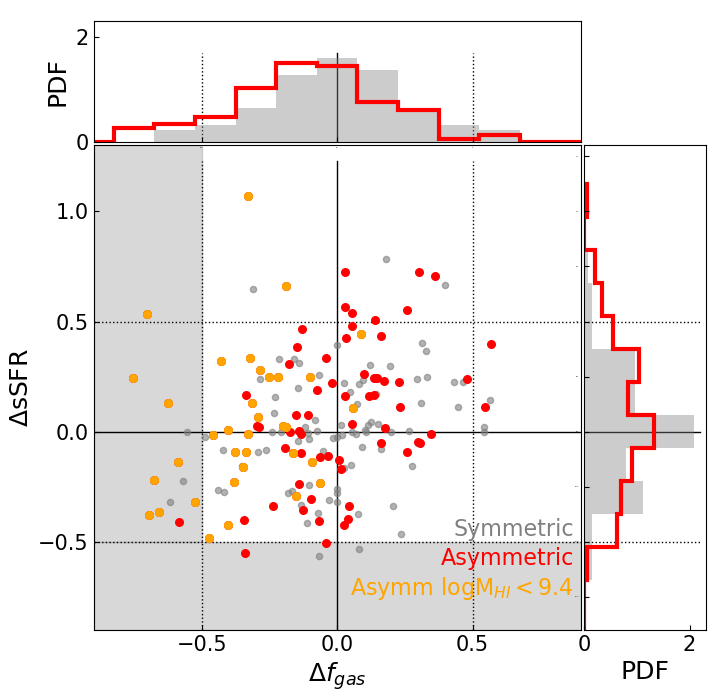}
    \caption{The $\dfg - \dsSFR$ parameter space as in Fig. \ref{fig:delta_AA},  but for xGASS$_{\rm MS}$. Asymmetric galaxies with \HI\ masses below $\lgMHI=9.4$, defined using the grey, dashed line in Fig. \ref{fig:samples} are highlighted with orange points.}
    \label{fig:delta_xG}
\end{figure}

{The results presented in the previous section may appear in contradiction with the recent work by \ct{watts20a} who, using xGASS, found that \HI\ asymmetries are preferentially found in gas-poor galaxies.}
{Here, we discuss how these two results can be easily reconciled.}
{In Fig. \ref{fig:delta_xG}, we show the  $\dfg-\dsSFR$ parameter space for xGASS$_{\rm MS}$.}
{As in ALFALFA$_{\rm MS}$, there are some galaxies with $\dsSFR>0.5$, though the median offset $\dsSFR = 0.03 \pm 0.04$ with $p=0.22$ is small, and we cannot determine that the two distributions are different.}
{The most significant difference compared to ALFALFA$_{\rm MS}$ is that the asymmetric population preferentially inhabits the $\dfg<0$ half of the parameter space, and there are now clearly galaxies with $\dfg<-0.5$.}
This is evident in the $\dfg$ distribution, which has a median $\dfg=-0.13 \pm 0.03$ and $p=0.03$ suggesting {that the asymmetric galaxies have a different distribution from the symmetric ones.} 
{This reflects the sensitive \HI\ observations of xGASS and the fact that a SFMS-selected sample does not translate to a clear selection in \HI\ mass fraction.}
{We demonstrate this by highlighting galaxies below an ALFALFA-like \HI\ mass limit of $\lgMHI=9.4$ (shown in the \HI\ mass fraction scaling relation panel of Fig. \ref{fig:samples} with a grey, dashed line) with orange points in Fig. \ref{fig:delta_xG}.}
{These galaxies constitute all but one of the $\dfg<-0.5$ points, and all apart from two have $\dfg<0$, showing that the `gas-poor' tail of the SFMS drives the suppressed \HI\ content of asymmetric galaxies. }
{Thus, in terms of their gas content, asymmetric galaxies preferentially scatter toward lower \HI\ fractions, but evidencing this requires deep \HI\ observations that are not typically reached by blind \HI\ surveys.}
{The xGASS$_{\rm MS}$ sample is consistent with ALFALFA$_{\rm MS}$ once selection biases are properly taken into account.}


\section{Discussion and conclusions}  \label{sec:concl}
{In this work, we have investigated the shape of global \HI\ spectra, through measurement of their asymmetry, in a data-set that samples the gas-rich regime with high statistics, covers $\sim2$ dex in \HI\ mass-fraction, and provides insight into strong gravitational interactions.}
{In particular, we have determined whether elevated global \HI\ asymmetry in galaxies is also associated with a change in their \HI\ content or sSFR.}
Our key result is that at fixed stellar mass, in the gas-rich regime, and on the SFMS, there is no systematic difference in the \HI\ content of galaxies with asymmetric global \HI\ spectra compared to symmetric ones. 
It is only mechanisms that disturb and remove the \HI\ that cause a clear, systematic, difference in the \HI\ content of  \HI\ asymmetric  galaxies, and this requires observations that measure \HI\ fractions below the median scaling relations {of representative samples}.
{There is some evidence for elevated sSFR in \HI\ asymmetric  galaxies, and this is primarily driven by galaxies with the highest \HI\ asymmetry.}
{There is also no clear correlation between optical morphology and global \HI\ asymmetry, and post-merger galaxies, despite being strongly disturbed systems optically, show a range of \HI\ asymmetries.}

Gas accretion has been proposed as a mechanism that could drive the ubiquity of disturbances in the \HI\ in galaxies. 
\ct{matthews98} found that 77\% of late-type galaxies had asymmetric global \HI\ spectra, while \ct{haynes98} found 45\% in isolated galaxies in general.
As late-type galaxies are typically more \HI\ rich, this has been interpreted as the signature of accretion-driven asymmetries. 
We find no evidence for an elevation in the \HI\ content of galaxies with asymmetric \HI\ spectra {in the gas-rich regime}, in agreement with \ct{reynolds20b}.
{The difference in these results} could be explained by the smaller sample sizes and the lack of compensation for measurement noise which \emph{must} be taken into account when comparing asymmetry rates. 
57\% of our ALFALFA spectra have $\Afr\geq1.1$, but after accounting for noise with the asymmetric population (before SFMS selection), i.e. 80\% confidence that their $\Afr\geq1.1$, this is reduced to 29\%. 
{To the same threshold, xGASS and xGASS isolated galaxies show asymmetry rates of 37\% and 32\%,  respectively.}
{Thus, gas-rich galaxies do not show a higher rate of \HI\ asymmetry than a sample that is more representative of the \HI\ fractions of galaxies, even when restricted to isolated systems.}
Assuming that gas-richness is a consequence of gas accretion, this implies that either gas accretion is not a dominant driver of global \HI\ asymmetry or that, if it is, it affects more the dynamics of the \HI\ reservoir (causing asymmetry in global spectra) than its \HI\ content.
Simulations of gas accretion on the galaxy scale have shown that, by mass, the dominant channel of gas accretion onto galaxies is through the accretion of satellite galaxies \cp{nelson13,nelson15}. 
This scenario brings additional gravitational perturbation to the picture, and the repeated merging of these satellites has been proposed to explain the ubiquity of disturbances in galaxies \cp{zaritsky97}.
{\ct{bassett17} showed that gas accreted onto a gas-rich galaxy will settle into co-rotating orbits, causing a smaller perturbation, whereas accretion onto a gas-poor galaxy allows the gas to retain memory of its accretion history, potentially causing greater asymmetry.}
Whether it is the active accretion of satellites, particularly those below the detection limit of many optical surveys \cp[e.g.][]{portas11,ramirez18}, the ongoing response to perturbations in the gravitational potential of the dark matter halo caused by these accretions \cp[e.g.][]{vaneymeren11b}, or a combination of both \cp{zaritsky13}, remain unknown. 
{Unfortunately, the marginal increase in sSFR that we have observed in galaxies with the strongest asymmetry cannot discriminate between these two scenarios, as both can cause increased sSFR \cp[e.g.][]{jog97,cox08}.}

Current cosmological-hydrodynamical simulations of galaxy formation are well suited to investigating \HI\ asymmetry in the gas-rich regime \cp[e.g.][]{watts20b}, and the relationship between gas accretion, tidal interactions, and global \HI\ asymmetry in these cosmological volumes remain to be quantified. 
Furthermore, {determining how disturbances move across the stellar, ionised, atomic, and molecular gas components of galaxies is an exciting prospect for future multi-wavelength observations.}

\section*{Acknowledgements}
ABW acknowledges the support of an Australian Government Research Training Program (RTP) Scholarship throughout the course of this work. LC is the recipient of an Australian Research Council Future Fellowship (FT180100066) funded by the Australian Government. Parts of this research were supported by the Australian Research Council Centre of Excellence for All-Sky Astrophysics in 3 Dimensions (ASTRO 3D), through project number CE170100013.

\section*{Data Availability}
The data that support the findings of this study are available upon request from the corresponding author, ABW.



\bibliographystyle{mnras}
\bibliography{bibfile} 

\newcommand{\noop}[1]{}
\begin{thebibliography}{}
\makeatletter
\relax
\def\mn@urlcharsother{\let\do\@makeother \do\$\do\&\do\#\do\^\do\_\do\%\do\~}
\def\mn@doi{\begingroup\mn@urlcharsother \@ifnextchar [ {\mn@doi@}
  {\mn@doi@[]}}
\def\mn@doi@[#1]#2{\def\@tempa{#1}\ifx\@tempa\@empty \href
  {http://dx.doi.org/#2} {doi:#2}\else \href {http://dx.doi.org/#2} {#1}\fi
  \endgroup}
\def\mn@eprint#1#2{\mn@eprint@#1:#2::\@nil}
\def\mn@eprint@arXiv#1{\href {http://arxiv.org/abs/#1} {{\tt arXiv:#1}}}
\def\mn@eprint@dblp#1{\href {http://dblp.uni-trier.de/rec/bibtex/#1.xml}
  {dblp:#1}}
\def\mn@eprint@#1:#2:#3:#4\@nil{\def\@tempa {#1}\def\@tempb {#2}\def\@tempc
  {#3}\ifx \@tempc \@empty \let \@tempc \@tempb \let \@tempb \@tempa \fi \ifx
  \@tempb \@empty \def\@tempb {arXiv}\fi \@ifundefined
  {mn@eprint@\@tempb}{\@tempb:\@tempc}{\expandafter \expandafter \csname
  mn@eprint@\@tempb\endcsname \expandafter{\@tempc}}}

\bibitem[\protect\citeauthoryear{{Abazajian} et~al.,}{{Abazajian}
  et~al.}{2009}]{abazajian09}
{Abazajian} K.~N.,  et~al., 2009, \mn@doi [ApJS] {10.1088/0067-0049/182/2/543},
  \href {https://ui.adsabs.harvard.edu/abs/2009ApJS..182..543A} {182, 543}

\bibitem[\protect\citeauthoryear{{Barnes} et~al.,}{{Barnes}
  et~al.}{2001}]{barnes01}
{Barnes} D.~G.,  et~al., 2001, \mn@doi [MNRAS]
  {10.1046/j.1365-8711.2001.04102.x}, \href
  {https://ui.adsabs.harvard.edu/abs/2001MNRAS.322..486B} {322, 486}

\bibitem[\protect\citeauthoryear{{Bassett}, {Bekki}, {Cortese}  \&
  {Couch}}{{Bassett} et~al.}{2017}]{bassett17}
{Bassett} R.,  {Bekki} K.,  {Cortese} L.,   {Couch} W.,  2017, \mn@doi [MNRAS]
  {10.1093/mnras/stx958}, \href
  {https://ui.adsabs.harvard.edu/abs/2017MNRAS.471.1892B} {471, 1892}

\bibitem[\protect\citeauthoryear{{Bok}, {Blyth}, {Gilbank}  \& {Elson}}{{Bok}
  et~al.}{2019}]{bok19}
{Bok} J.,  {Blyth} S.-L.,  {Gilbank} D.~G.,   {Elson} E.~C.,  2019, \mn@doi
  [MNRAS] {10.1093/mnras/sty3448}, \href
  {http://adsabs.harvard.edu/abs/2019MNRAS.484..582B} {484, 582}

\bibitem[\protect\citeauthoryear{{Catinella} et~al.,}{{Catinella}
  et~al.}{2018}]{catinella18}
{Catinella} B.,  et~al., 2018, \mn@doi [MNRAS] {10.1093/mnras/sty089}, \href
  {https://ui.adsabs.harvard.edu/#abs/2018MNRAS.476..875C} {476, 875}

\bibitem[\protect\citeauthoryear{{Chung}, {van Gorkom}, {Kenney}  \&
  {Vollmer}}{{Chung} et~al.}{2007}]{chung07}
{Chung} A.,  {van Gorkom} J.~H.,  {Kenney} J. D.~P.,   {Vollmer} B.,  2007,
  \mn@doi [ApJL] {10.1086/518034}, \href
  {https://ui.adsabs.harvard.edu/abs/2007ApJ...659L.115C} {659, L115}

\bibitem[\protect\citeauthoryear{{Cortese}, {Catinella}, {Cook}  \&
  {Janowiecki}}{{Cortese} et~al.}{2020}]{cortese20}
{Cortese} L.,  {Catinella} B.,  {Cook} R.~H.~W.,   {Janowiecki} S.,  2020,
  \mn@doi [MNRAS] {10.1093/mnrasl/slaa032}, \href
  {https://ui.adsabs.harvard.edu/abs/2020MNRAS.494L..42C} {494, L42}

\bibitem[\protect\citeauthoryear{{Cox}, {Jonsson}, {Somerville}, {Primack}  \&
  {Dekel}}{{Cox} et~al.}{2008}]{cox08}
{Cox} T.~J.,  {Jonsson} P.,  {Somerville} R.~S.,  {Primack} J.~R.,   {Dekel}
  A.,  2008, \mn@doi [MNRAS] {10.1111/j.1365-2966.2007.12730.x}, \href
  {https://ui.adsabs.harvard.edu/abs/2008MNRAS.384..386C} {384, 386}

\bibitem[\protect\citeauthoryear{{Darg} et~al.,}{{Darg} et~al.}{2010}]{darg10}
{Darg} D.~W.,  et~al., 2010, \mn@doi [MNRAS]
  {10.1111/j.1365-2966.2009.15786.x}, \href
  {https://ui.adsabs.harvard.edu/abs/2010MNRAS.401.1552D} {401, 1552}

\bibitem[\protect\citeauthoryear{{Deg}, {Blyth}, {Hank}, {Kruger}  \&
  {Carignan}}{{Deg} et~al.}{2020}]{deg20}
{Deg} N.,  {Blyth} S.~L.,  {Hank} N.,  {Kruger} S.,   {Carignan} C.,  2020,
  \mn@doi [MNRAS] {10.1093/mnras/staa1368}, \href
  {https://ui.adsabs.harvard.edu/abs/2020MNRAS.495.1984D} {495, 1984}

\bibitem[\protect\citeauthoryear{{Ellison}, {Mendel}, {Patton}  \&
  {Scudder}}{{Ellison} et~al.}{2013}]{ellison13}
{Ellison} S.~L.,  {Mendel} J.~T.,  {Patton} D.~R.,   {Scudder} J.~M.,  2013,
  \mn@doi [MNRAS] {10.1093/mnras/stt1562}, \href
  {https://ui.adsabs.harvard.edu/abs/2013MNRAS.435.3627E} {435, 3627}

\bibitem[\protect\citeauthoryear{{Ellison}, {Catinella}  \&
  {Cortese}}{{Ellison} et~al.}{2018}]{ellison18}
{Ellison} S.~L.,  {Catinella} B.,   {Cortese} L.,  2018, \mn@doi [MNRAS]
  {10.1093/mnras/sty1247}, \href
  {https://ui.adsabs.harvard.edu/#abs/2018MNRAS.478.3447E} {478, 3447}

\bibitem[\protect\citeauthoryear{{Espada}, {Verdes-Montenegro}, {Huchtmeier},
  {Sulentic}, {Verley}, {Leon}  \& {Sabater}}{{Espada} et~al.}{2011}]{espada11}
{Espada} D.,  {Verdes-Montenegro} L.,  {Huchtmeier} W.~K.,  {Sulentic} J.,
  {Verley} S.,  {Leon} S.,   {Sabater} J.,  2011, \mn@doi [A\&A]
  {10.1051/0004-6361/201016117}, \href
  {https://ui.adsabs.harvard.edu/#abs/2011A&A...532A.117E} {532, A117}

\bibitem[\protect\citeauthoryear{{Feng}, {Shen}, {Yuan}, {Riffel}  \&
  {Pan}}{{Feng} et~al.}{2020}]{feng20}
{Feng} S.,  {Shen} S.-Y.,  {Yuan} F.-T.,  {Riffel} R.~A.,   {Pan} K.,  2020,
  \mn@doi [ApJL] {10.3847/2041-8213/ab7dba}, \href
  {https://ui.adsabs.harvard.edu/abs/2020ApJ...892L..20F} {892, L20}

\bibitem[\protect\citeauthoryear{{Giovanelli} et~al.,}{{Giovanelli}
  et~al.}{2005}]{giovanelli05}
{Giovanelli} R.,  et~al., 2005, \mn@doi [AJ] {10.1086/497431}, \href
  {https://ui.adsabs.harvard.edu/#abs/2005AJ....130.2598G} {130, 2598}

\bibitem[\protect\citeauthoryear{{Haynes}, {Hogg}, {Maddalena}, {Roberts}  \&
  {van Zee}}{{Haynes} et~al.}{1998}]{haynes98}
{Haynes} M.~P.,  {Hogg} D.~E.,  {Maddalena} R.~J.,  {Roberts} M.~S.,   {van
  Zee} L.,  1998, \mn@doi [AJ] {10.1086/300166}, \href
  {https://ui.adsabs.harvard.edu/#abs/1998AJ....115...62H} {115, 62}

\bibitem[\protect\citeauthoryear{{Haynes} et~al.,}{{Haynes}
  et~al.}{2018}]{haynes18}
{Haynes} M.~P.,  et~al., 2018, \mn@doi [ApJ] {10.3847/1538-4357/aac956}, \href
  {http://adsabs.harvard.edu/abs/2018ApJ...861...49H} {861, 49}

\bibitem[\protect\citeauthoryear{{Janowiecki}, {Catinella}, {Cortese},
  {Saintonge}, {Brown}  \& {Wang}}{{Janowiecki} et~al.}{2017}]{janowiecki17}
{Janowiecki} S.,  {Catinella} B.,  {Cortese} L.,  {Saintonge} A.,  {Brown} T.,
   {Wang} J.,  2017, \mn@doi [MNRAS] {10.1093/mnras/stx046}, \href
  {https://ui.adsabs.harvard.edu/abs/2017MNRAS.466.4795J} {466, 4795}

\bibitem[\protect\citeauthoryear{{Janowiecki}, {Catinella}, {Cortese},
  {Saintonge}  \& {Wang}}{{Janowiecki} et~al.}{2020}]{janowiecki20}
{Janowiecki} S.,  {Catinella} B.,  {Cortese} L.,  {Saintonge} A.,   {Wang} J.,
  2020, \mn@doi [MNRAS] {10.1093/mnras/staa178}, \href
  {https://ui.adsabs.harvard.edu/abs/2020MNRAS.493.1982J} {493, 1982}

\bibitem[\protect\citeauthoryear{{Jog}}{{Jog}}{1997}]{jog97}
{Jog} C.~J.,  1997, \mn@doi [ApJ] {10.1086/304721}, \href
  {https://ui.adsabs.harvard.edu/abs/1997ApJ...488..642J} {488, 642}

\bibitem[\protect\citeauthoryear{{Jog} \& {Combes}}{{Jog} \&
  {Combes}}{2009}]{jog09}
{Jog} C.~J.,  {Combes} F.,  2009, \mn@doi [Phys.Rep.]
  {10.1016/j.physrep.2008.12.002}, \href
  {https://ui.adsabs.harvard.edu/#abs/2009PhR...471...75J} {471, 75}

\bibitem[\protect\citeauthoryear{{Koribalski} et~al.,}{{Koribalski}
  et~al.}{2020}]{koribalski20}
{Koribalski} B.~S.,  et~al., 2020, \mn@doi [ApSS] {10.1007/s10509-020-03831-4},
  \href {https://ui.adsabs.harvard.edu/abs/2020Ap&SS.365..118K} {365, 118}

\bibitem[\protect\citeauthoryear{{Manthey}, {Aalto}, {H{\"u}ttemeister}  \&
  {Oosterloo}}{{Manthey} et~al.}{2008}]{manthey08}
{Manthey} E.,  {Aalto} S.,  {H{\"u}ttemeister} S.,   {Oosterloo} T.~A.,  2008,
  \mn@doi [A\&A] {10.1051/0004-6361:20077584}, \href
  {https://ui.adsabs.harvard.edu/abs/2008A&A...484..693M} {484, 693}

\bibitem[\protect\citeauthoryear{{Martin} et~al.,}{{Martin}
  et~al.}{2005}]{martin05}
{Martin} D.~C.,  et~al., 2005, \mn@doi [ApJL] {10.1086/426387}, \href
  {https://ui.adsabs.harvard.edu/abs/2005ApJ...619L...1M} {619, L1}

\bibitem[\protect\citeauthoryear{{Matthews}, {van Driel}  \&
  {Gallagher}}{{Matthews} et~al.}{1998}]{matthews98}
{Matthews} L.~D.,  {van Driel} W.,   {Gallagher} J.~S. I.,  1998, \mn@doi [AJ]
  {10.1086/300492}, \href
  {https://ui.adsabs.harvard.edu/#abs/1998AJ....116.1169M} {116, 1169}

\bibitem[\protect\citeauthoryear{{Meyer} et~al.,}{{Meyer}
  et~al.}{2004}]{meyer04}
{Meyer} M.~J.,  et~al., 2004, \mn@doi [MNRAS]
  {10.1111/j.1365-2966.2004.07710.x}, \href
  {https://ui.adsabs.harvard.edu/abs/2004MNRAS.350.1195M} {350, 1195}

\bibitem[\protect\citeauthoryear{{Nair} \& {Abraham}}{{Nair} \&
  {Abraham}}{2010}]{nair10}
{Nair} P.~B.,  {Abraham} R.~G.,  2010, \mn@doi [ApJS]
  {10.1088/0067-0049/186/2/427}, \href
  {https://ui.adsabs.harvard.edu/abs/2010ApJS..186..427N} {186, 427}

\bibitem[\protect\citeauthoryear{{Nelson}, {Vogelsberger}, {Genel}, {Sijacki},
  {Kere{\v{s}}}, {Springel}  \& {Hernquist}}{{Nelson} et~al.}{2013}]{nelson13}
{Nelson} D.,  {Vogelsberger} M.,  {Genel} S.,  {Sijacki} D.,  {Kere{\v{s}}} D.,
   {Springel} V.,   {Hernquist} L.,  2013, \mn@doi [MNRAS]
  {10.1093/mnras/sts595}, \href
  {https://ui.adsabs.harvard.edu/abs/2013MNRAS.429.3353N} {429, 3353}

\bibitem[\protect\citeauthoryear{{Nelson}, {Genel}, {Vogelsberger}, {Springel},
  {Sijacki}, {Torrey}  \& {Hernquist}}{{Nelson} et~al.}{2015}]{nelson15}
{Nelson} D.,  {Genel} S.,  {Vogelsberger} M.,  {Springel} V.,  {Sijacki} D.,
  {Torrey} P.,   {Hernquist} L.,  2015, \mn@doi [MNRAS] {10.1093/mnras/stv017},
  \href {https://ui.adsabs.harvard.edu/abs/2015MNRAS.448...59N} {448, 59}

\bibitem[\protect\citeauthoryear{{Nevin} et~al.,}{{Nevin}
  et~al.}{2021}]{nevin21}
{Nevin} R.,  et~al., 2021, arXiv e-prints, \href
  {https://ui.adsabs.harvard.edu/abs/2021arXiv210202208N} {p. arXiv:2102.02208}

\bibitem[\protect\citeauthoryear{{Portas} et~al.,}{{Portas}
  et~al.}{2011}]{portas11}
{Portas} A.,  et~al., 2011, \mn@doi [ApJ] {10.1088/2041-8205/739/1/L27}, \href
  {https://ui.adsabs.harvard.edu/abs/2011ApJ...739L..27P} {739, L27}

\bibitem[\protect\citeauthoryear{{Ram{\'\i}rez-Moreta}
  et~al.,}{{Ram{\'\i}rez-Moreta} et~al.}{2018}]{ramirez18}
{Ram{\'\i}rez-Moreta} P.,  et~al., 2018, \mn@doi [A\&A]
  {10.1051/0004-6361/201833333}, \href
  {https://ui.adsabs.harvard.edu/abs/2018A&A...619A.163R} {619, A163}

\bibitem[\protect\citeauthoryear{{Reynolds}, {Westmeier}, {Staveley-Smith},
  {Chauhan}  \& {Lagos}}{{Reynolds} et~al.}{2020a}]{reynolds20a}
{Reynolds} T.~N.,  {Westmeier} T.,  {Staveley-Smith} L.,  {Chauhan} G.,
  {Lagos} C.~D.~P.,  2020a, \mn@doi [MNRAS] {10.1093/mnras/staa597}, \href
  {https://ui.adsabs.harvard.edu/abs/2020MNRAS.tmp..562R} {}

\bibitem[\protect\citeauthoryear{{Reynolds}, {Westmeier}  \&
  {Staveley-Smith}}{{Reynolds} et~al.}{2020b}]{reynolds20b}
{Reynolds} T.~N.,  {Westmeier} T.,   {Staveley-Smith} L.,  2020b, arXiv
  e-prints, \href {https://ui.adsabs.harvard.edu/abs/2020arXiv201003720R} {p.
  arXiv:2010.03720}

\bibitem[\protect\citeauthoryear{{Richter} \& {Sancisi}}{{Richter} \&
  {Sancisi}}{1994}]{richter94}
{Richter} O.~G.,  {Sancisi} R.,  1994, A\&A, \href
  {https://ui.adsabs.harvard.edu/#abs/1994A&A...290L...9R} {290, L9}

\bibitem[\protect\citeauthoryear{{Saintonge}}{{Saintonge}}{2007}]{saintonge07}
{Saintonge} A.,  2007, \mn@doi [AJ] {10.1086/513515}, \href
  {https://ui.adsabs.harvard.edu/#abs/2007AJ....133.2087S} {133, 2087}

\bibitem[\protect\citeauthoryear{{Salim} et~al.,}{{Salim}
  et~al.}{2016}]{salim16}
{Salim} S.,  et~al., 2016, \mn@doi [ApJS] {10.3847/0067-0049/227/1/2}, \href
  {https://ui.adsabs.harvard.edu/abs/2016ApJS..227....2S} {227, 2}

\bibitem[\protect\citeauthoryear{{Salim}, {Boquien}  \& {Lee}}{{Salim}
  et~al.}{2018}]{salim18}
{Salim} S.,  {Boquien} M.,   {Lee} J.~C.,  2018, \mn@doi [ApJS]
  {10.3847/1538-4357/aabf3c}, \href
  {https://ui.adsabs.harvard.edu/abs/2018ApJ...859...11S} {859, 11}

\bibitem[\protect\citeauthoryear{{Sancisi}, {Fraternali}, {Oosterloo}  \& {van
  der Hulst}}{{Sancisi} et~al.}{2008}]{sancisi08}
{Sancisi} R.,  {Fraternali} F.,  {Oosterloo} T.,   {van der Hulst} T.,  2008,
  \mn@doi [A\&AR] {10.1007/s00159-008-0010-0}, \href
  {https://ui.adsabs.harvard.edu/#abs/2008A&ARv..15..189S} {15, 189}

\bibitem[\protect\citeauthoryear{{Schiminovich}, {van Gorkom}  \& {van der
  Hulst}}{{Schiminovich} et~al.}{2013}]{schiminovich13}
{Schiminovich} D.,  {van Gorkom} J.~H.,   {van der Hulst} J.~M.,  2013, \mn@doi
  [AJ] {10.1088/0004-6256/145/2/34}, \href
  {https://ui.adsabs.harvard.edu/abs/2013AJ....145...34S} {145, 34}

\bibitem[\protect\citeauthoryear{{Simard}, {Mendel}, {Patton}, {Ellison}  \&
  {McConnachie}}{{Simard} et~al.}{2011}]{simard11}
{Simard} L.,  {Mendel} J.~T.,  {Patton} D.~R.,  {Ellison} S.~L.,
  {McConnachie} A.~W.,  2011, \mn@doi [ApJS] {10.1088/0067-0049/196/1/11},
  \href {https://ui.adsabs.harvard.edu/abs/2011ApJS..196...11S} {196, 11}

\bibitem[\protect\citeauthoryear{{Swaters}, {Schoenmakers}, {Sancisi}  \& {van
  Albada}}{{Swaters} et~al.}{1999}]{swaters99}
{Swaters} R.~A.,  {Schoenmakers} R.~H.~M.,  {Sancisi} R.,   {van Albada} T.~S.,
   1999, \mn@doi [MNRAS] {10.1046/j.1365-8711.1999.02332.x}, \href
  {https://ui.adsabs.harvard.edu/#abs/1999MNRAS.304..330S} {304, 330}

\bibitem[\protect\citeauthoryear{{Walter}, {Brinks}, {de Blok}, {Bigiel},
  {Kennicutt}, {Thornley}  \& {Leroy}}{{Walter} et~al.}{2008}]{walter08}
{Walter} F.,  {Brinks} E.,  {de Blok} W.~J.~G.,  {Bigiel} F.,  {Kennicutt}
  Robert~C. J.,  {Thornley} M.~D.,   {Leroy} A.,  2008, \mn@doi [AJ]
  {10.1088/0004-6256/136/6/2563}, \href
  {https://ui.adsabs.harvard.edu/#abs/2008AJ....136.2563W} {136, 2563}

\bibitem[\protect\citeauthoryear{{Wang} et~al.,}{{Wang} et~al.}{2014}]{wang14}
{Wang} J.,  et~al., 2014, \mn@doi [MNRAS] {10.1093/mnras/stu649}, \href
  {https://ui.adsabs.harvard.edu/#abs/2014MNRAS.441.2159W} {441, 2159}

\bibitem[\protect\citeauthoryear{{Watts}, {Catinella}, {Cortese}  \&
  {Power}}{{Watts} et~al.}{2020a}]{watts20a}
{Watts} A.~B.,  {Catinella} B.,  {Cortese} L.,   {Power} C.,  2020a, \mn@doi
  [MNRAS] {10.1093/mnras/staa094}, \href
  {https://ui.adsabs.harvard.edu/abs/2020MNRAS.492.3672W} {492, 3672}

\bibitem[\protect\citeauthoryear{{Watts}, {Power}, {Catinella}, {Cortese}  \&
  {Stevens}}{{Watts} et~al.}{2020b}]{watts20b}
{Watts} A.~B.,  {Power} C.,  {Catinella} B.,  {Cortese} L.,   {Stevens} A.
  R.~H.,  2020b, \mn@doi [MNRAS] {10.1093/mnras/staa3200}, \href
  {https://ui.adsabs.harvard.edu/abs/2020MNRAS.499.5205W} {499, 5205}

\bibitem[\protect\citeauthoryear{{Westmeier}, {Jurek}, {Obreschkow},
  {Koribalski}  \& {Staveley-Smith}}{{Westmeier} et~al.}{2014}]{westmeier14}
{Westmeier} T.,  {Jurek} R.,  {Obreschkow} D.,  {Koribalski} B.~S.,
  {Staveley-Smith} L.,  2014, \mn@doi [MNRAS] {10.1093/mnras/stt2266}, \href
  {https://ui.adsabs.harvard.edu/#abs/2014MNRAS.438.1176W} {438, 1176}

\bibitem[\protect\citeauthoryear{{Wong} et~al.,}{{Wong} et~al.}{2006}]{wong06}
{Wong} O.~I.,  et~al., 2006, \mn@doi [MNRAS]
  {10.1111/j.1365-2966.2006.10846.x}, \href
  {https://ui.adsabs.harvard.edu/abs/2006MNRAS.371.1855W} {371, 1855}

\bibitem[\protect\citeauthoryear{{Wright} et~al.,}{{Wright}
  et~al.}{2010}]{wright10}
{Wright} E.~L.,  et~al., 2010, \mn@doi [AJ] {10.1088/0004-6256/140/6/1868},
  \href {https://ui.adsabs.harvard.edu/abs/2010AJ....140.1868W} {140, 1868}

\bibitem[\protect\citeauthoryear{{York} et~al.,}{{York} et~al.}{2000}]{york00}
{York} D.~G.,  et~al., 2000, \mn@doi [AJ] {10.1086/301513}, \href
  {https://ui.adsabs.harvard.edu/abs/2000AJ....120.1579Y} {120, 1579}

\bibitem[\protect\citeauthoryear{{Zaritsky} \& {Rix}}{{Zaritsky} \&
  {Rix}}{1997}]{zaritsky97}
{Zaritsky} D.,  {Rix} H.-W.,  1997, \mn@doi [ApJ] {10.1086/303692}, \href
  {https://ui.adsabs.harvard.edu/#abs/1997ApJ...477..118Z} {477, 118}

\bibitem[\protect\citeauthoryear{{Zaritsky} et~al.,}{{Zaritsky}
  et~al.}{2013}]{zaritsky13}
{Zaritsky} D.,  et~al., 2013, \mn@doi [ApJ] {10.1088/0004-637X/772/2/135},
  \href {https://ui.adsabs.harvard.edu/abs/2013ApJ...772..135Z} {772, 135}

\bibitem[\protect\citeauthoryear{{Zu} \& {Mandelbaum}}{{Zu} \&
  {Mandelbaum}}{2016}]{zu16}
{Zu} Y.,  {Mandelbaum} R.,  2016, \mn@doi [MNRAS] {10.1093/mnras/stw221}, \href
  {https://ui.adsabs.harvard.edu/abs/2016MNRAS.457.4360Z} {457, 4360}

\bibitem[\protect\citeauthoryear{{de Blok} et~al.,}{{de Blok}
  et~al.}{2020}]{deblok20}
{de Blok} W.~J.~G.,  et~al., 2020, arXiv e-prints, \href
  {https://ui.adsabs.harvard.edu/abs/2020arXiv200909766D} {p. arXiv:2009.09766}

\bibitem[\protect\citeauthoryear{{van Eymeren}, {J{\"u}tte}, {Jog}, {Stein}  \&
  {Dettmar}}{{van Eymeren} et~al.}{2011}]{vaneymeren11b}
{van Eymeren} J.,  {J{\"u}tte} E.,  {Jog} C.~J.,  {Stein} Y.,   {Dettmar}
  R.~J.,  2011, \mn@doi [A\&A] {10.1051/0004-6361/201016178}, \href
  {https://ui.adsabs.harvard.edu/abs/2011A&A...530A..30V} {530, A30}

\makeatother
\end{thebibliography}




\bsp	
\label{lastpage}
\end{document}